\documentclass[reprint,superscriptaddress,
amsmath,amssymb,
aps,
pra
]{revtex4-2}

\usepackage{graphicx}
\graphicspath{{Images/}}
\usepackage{xcolor}
\usepackage[colorlinks=true,citecolor=blue,urlcolor=blue]{hyperref}
\usepackage{mathrsfs}   % for math script letters
\usepackage{amsthm}     % for theorems, definitions, and lemmas
\usepackage{amsmath}    % for \begin{split}
\usepackage{amssymb}
\usepackage{physics}    % for \Tr command

\usepackage{float}
\usepackage[caption=false]{subfig}

\usepackage{array}
\usepackage{multirow}
\usepackage{booktabs}   % for spacings between table cols
\newtheorem{thm}{Theorem}

\newtheorem{claim}{Claim}

\newcommand{\vars}[1]{\mathcal{#1}}
\newcommand{\Var}{\mathrm{Var}}
\newcommand{\sX}{\mathsf{X}}
\newcommand{\sZ}{\mathsf{Z}}
\begin{document}

\title{Sampled sub-block hashing for large input randomness extraction}

\author{Hong Jie Ng}
\thanks{These two authors contributed equally}
\affiliation{Department of Electrical \& Computer Engineering, National University of Singapore, Singapore }

\author{Wen Yu Kon}
\thanks{These two authors contributed equally}
\affiliation{Department of Electrical \& Computer Engineering, National University of Singapore, Singapore }

\author{Ignatius William Primaatmaja}
\affiliation{Department of Electrical \& Computer Engineering, National University of Singapore, Singapore }
\affiliation{Squareroot8 Technologies Pte Ltd, Singapore}

\author{Chao Wang}
\affiliation{Department of Electrical \& Computer Engineering, National University of Singapore, Singapore }

\author{Charles Lim}
\email{charleslim.research@gmail.com}
\affiliation{Department of Electrical \& Computer Engineering, National University of Singapore, Singapore }

\begin{abstract}
Randomness extraction is an essential post-processing step in practical quantum cryptography systems. When statistical fluctuations are taken into consideration, the requirement of large input data size could heavily penalise the speed and resource consumption of the randomness extraction process, thereby limiting the overall system performance. In this work, we propose a sampled sub-block hashing approach to circumvent this problem by randomly dividing the large input block into multiple sub-blocks and processing them individually. Through simulations and experiments, we demonstrate that our method achieves an order-of-magnitude improvement in system throughput while keeping the resource utilisation low. Furthermore, our proposed approach is applicable to a generic class of quantum cryptographic protocols that satisfy the generalised entropy accumulation framework, presenting a highly promising and general solution for high-speed post-processing in quantum cryptographic applications such as quantum key distribution and quantum random number generation. 
\end{abstract}

\maketitle

\section{Introduction} \label{sec:intro}

Randomness extractors are the cornerstone of quantum cryptography. It is an essential step in both quantum key distribution (QKD)~\cite{bennett_quantum_1984, ekert_quantum_1991, bennett_quantum_1992} and quantum random number generation (QRNG)~\cite{stefanov_optical_2000}, which are two of the most mature quantum technologies to date. In general, after execution of a quantum cryptographic protocol, the raw data collected is a weakly random source~\cite{chor_unbiased_1988}, which the adversary, Eve, might have some side-information on. Randomness extractors, as their name suggests, are functions, with the help of a short random seed, that transform the weakly random source to an (almost) perfectly random source, even from Eve's perspective. 

Often, Toeplitz hashing~\cite{krawczyk_lfsr-based_1994} is the function chosen for randomness extraction. This is because Toeplitz hashing, which belongs to a family of universal hash functions~\cite{carter_universal_1979}, is optimal in terms of extractable length~\cite{impagliazzo_pseudo-random_1989, radhakrishnan_bounds_2000}. In addition, Toeplitz hashing also has a straightforward construction and implementation. Toeplitz hashing utilises a Toeplitz matrix $\mathbf{T}$, which is a diagonal-constant matrix, constructed using a uniform random seed. The number of rows and columns of $\mathbf{T}$ correspond to the size of the output and input of the extractor, respectively. Expressing the input data as a column vector $A'$, the entire extraction process can be simply expressed as $K = \mathbf{T} \cdot A'$, where $K$ is the extracted output.

Field programmable gate arrays (FPGAs) are an excellent choice of implementation platform for Toeplitz hashing because computations can be executed in parallel compared to sequentially on CPUs. In addition, FPGAs consume much less power~\cite{awad_fpga_2009} and are much smaller than CPUs. An FPGA implementation also eases the transition to application-specific integrated circuits (ASIC) compared to a CPU implementation. These factors are all crucial for future downstream engineering work to achieve chip-based implementations of QKD and QRNG systems, which are of keen interest to both academia and the industry. However, as FPGAs are platforms with limited resources, active steps have to be taken to keep the resource utilisation level low.

On high-level platforms like CPUs and GPUs, the process of matrix-vector multiplication can be sped-up using the fast Fourier transform (FFT) algorithm. While it is possible to implement FFT-based Toeplitz hashing on FPGA as well~\cite{li_high-speed_2019}, this is often not the chosen method, because FFT requires floating-point precision, which would drive up the complexity and resource consumption of the implementation. In addition, it is also hard to determine if FFT will improve the speed of Toeplitz hashing on FPGA since the architectures of CPU and FPGA are different. Hence, in this paper, we only consider a direct hashing approach implemented on FPGA. 

All real implementations of quantum cryptographic protocols are run for a finite duration, and therefore have finite resources. Due to statistical fluctuations in finite-size data, randomness extraction has to be performed on large input sizes for practical implementations of quantum cryptographic protocols~\cite{scarani_quantum_2008}. For example, in QKD, finite-size analysis requires a block size of at least $10^6$ bits~\cite{leverrier_finite-size_2010, lim_concise_2014}. When considering protocols with weaker assumptions, such as the measurement device-independent (MDI) setting~\cite{curty_finite-key_2014, nie_experimental_2016, wang_provably-secure_2023} and the device-independent (DI) setting~\cite{acin_device-independent_2007, pironio_device-independent_2009, liu_device-independent_2021}, the block size required is even larger, around $10^{11}$ bits. Taking DI-QKD as an example, despite improvements in both theoretical~\cite{schwonnek_device-independent_2021, tan_improved_2022} and experimental~\cite{zhang_device-independent_2022, nadlinger_experimental_2022} aspects, the required block size is still larger than the input sizes of state-of-the-art implementations of Toeplitz hashing on FPGA~\cite{constantin_fpga-based_2017, li_high-speed_2019, drahi_certified_2020}.

This is because the hashing process becomes increasingly impractical as the input size increases. This can be attributed to two scenarios, which are a decrease in throughput, or an increase in resources required. As the input size of hashing increases, the output speed naturally decreases, as a larger number of bits have to be processed to produce even a single secure bit. Alternatively, to prevent the throughput of hashing from decreasing, the FPGA can create duplicates of certain hardware and execute them in parallel. However, this leads to a large increase in the resource utilisation, which restricts the quantum cryptographic system to the use of high-end FPGAs.

To ensure that the hashing process remains practical while satisfying the finite-size effect, we propose a sampled sub-block hashing approach. The main idea is to randomly split (via sampling) the large input data into sub-blocks, and provide a lower bound on the conditional smooth min-entropy of each sub-block. With this bound, we then perform randomness extraction on each sub-block, and concatenate the outputs together. If our proposed method is performed on a protocol satisfying certain properties and whose security can be analysed with the recent generalised entropy accumulation theorem (GEAT), we can demonstrate that our method only introduces a small (linear) loss in security while maintaining the same number of secure bits obtained.

The sample-and-hash method by Konig and Renner~\cite{konig_sampling_2011}, and the sample-then-extract method by Dupuis et al.~\cite{dupuis_entanglement_2015} contains features that closely resemble those of our method. However, they incur high entropy losses. The reasons are two-fold. First, the sampling penalty of our method is lower because works~\cite{konig_sampling_2011, dupuis_entanglement_2015} have fewer assumptions about the characterisation of the entropy source. Thus, our method considers a more restricted class of protocols (but not too restrictive in general). Secondly, works~\cite{konig_sampling_2011, dupuis_entanglement_2015} also do not account for the correlation between different sampled sub-blocks. Therefore, after performing the sampling, the rest of the unsampled bits are discarded. The act of discarding the other bits effectively introduces a large loss of entropy. Using our method, we can account for the correlation between different sampled sub-blocks. Thus, we are able to perform hashing on all the sub-blocks and concatenate the outputs, while still preserving the overall security.

We employ our method on a simulated standard BBM92~\cite{bennett_quantum_1992} QKD protocol as a proof of concept demonstration. We show that with our method, we are able to reduce the execution time of Toeplitz hashing by almost twenty-fold. It should be noted that the improvement can be increased further by optimising the sampling parameters.

The rest of the paper is organised as follows. 
In section \ref{sec:framework}, we provide a description of the theoretical framework, GEAT, that we use in our analysis.
In section \ref{sec:theory}, the theoretical analysis for our method will be explained in detail. The implementation and results of our proposed method is given in section \ref{sec:results}. Finally, some discussions and a conclusion is given in section \ref{sec:conclusion}.

\section{Theoretical framework} \label{sec:framework}

\begin{figure}[!t]
    \centering
    \includegraphics[width=\columnwidth]{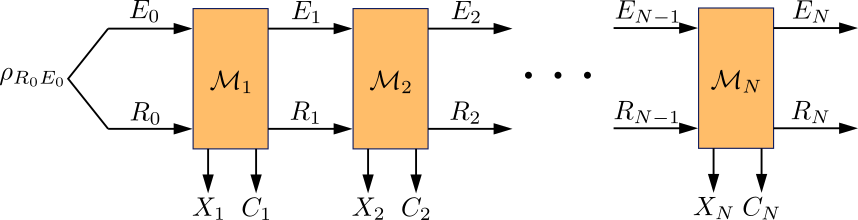}
    \caption{Sequential model for generalised EAT~\cite{Metger_EAT2022}, where an initial state $\rho_{R_0E_0}$ undergoes a series of EAT channels, $\vars{M}_i$, to generate the final state $\rho_{X^N C^NR_N E_N}$.} 
    
    \label{fig:GeneralisedEATModel}
\end{figure}

Our work makes use of the generalised entropy accumulation theorem~\cite{Metger_EAT2022}, and we provide a brief description of it here. 
Suppose at the end of some protocol, we have the output quantum state 
$\rho_{X^NC^NR_NE_N}$ which can be described as being generated by applying a sequence of channels $\vars{M}_1$, $\cdots$, $\vars{M}_N$ (termed EAT channels), $\vars{M}_i:R_{i-1}E_{i-1}\rightarrow X_iR_iC_iE_i$, on some initial state $\rho_{R_0E_0}$ (see Fig.~\ref{fig:GeneralisedEATModel}).
In the output state, $X^N$ typically refers to the output (e.g., raw keys in QKD), $C^N$ refer to the statistics used to select an event (e.g., whether the protocol should be aborted), $R_i$ are the side-information carried by Alice and Bob, while $E_i$ is Eve's side-information.
Generalised EAT provides a tight bound on the conditional smooth min-entropy of the output $X^N$ conditioned on the side-information $E_N$, i.e., $H_{\text{min}}^{\varepsilon}(X^N|E_N)_{\rho_{X^NC^NR_NE_N|\Omega}}$, where $\Omega\subset\vars{C}^N$ is a set of events. This bound is extremely useful as it is directly related to the final key length and the security parameter in quantum cryptographic protocols. 

Generalised EAT states that if the following two conditions:
\begin{enumerate}
    \item \emph{Non-signalling}: One requires that any side information that Eve can have about $X_1\cdots X_{i-1}$ should already be contained in $E_{i-1}$, and not stored in $R_{i-1}$ and revealed in later rounds (for instance to $E_i$). More formally, for each EAT channel $\vars{M}_i$, there exists a channel $\vars{R}_i:E_{i-1}\rightarrow E_i$, such that $\Tr_{X_iR_iC_i}\circ\vars{M}_i=\vars{R}_i\circ\Tr_{R_{i-1}}$,
    where $\circ$ represents the composition of the channels.
    \item \emph{Projective reconstructability}: Statistics $C^N$ should be able to be reconstructed by performing a projective measurement on $X^N$ and $E_N$, and applying a function on the outcomes. More formally, there exists a channel $\vars{T}$ such that $\vars{M}_N\circ\cdots\circ\vars{M}_1=\vars{T}\circ\Tr_{C_N}\circ\vars{M}_N\circ\cdots\circ\Tr_{C_1}\circ\vars{M}_1$.
\end{enumerate}
are satisfied, then (original form presented in Appendix~\ref{app:GEAT})
\begin{equation}
\label{eqn:GenEATEqn}
    H_{\text{min}}^{\varepsilon}(X^N|E_N)_{\rho_{X^NC^NR_NE_N|\Omega}}\geq Nh-v_1\sqrt{N}-v_0
\end{equation}
where
\begin{gather*}
    h=\min_{c^N\in\Omega}f(\text{freq}(c^N))\\
    v_0=\frac{\beta(\varepsilon)}{V^2} 2^{\frac{1-\xi}{\xi}\nu} \ln^3(2^{\nu}+e^2),\quad v_1=\gamma(\varepsilon) V\\
    \nu=2\log_2 d_X+\text{Max}(f)-\text{Min}(f)\\
    V=\log_2(2d_X^2+1)+\sqrt{2+\Var(f)}
\end{gather*}
with variables $\xi$, $\gamma$ and $\beta$, which are independent on $f$, and $\Var(f)$, $\text{Max}(f)$ and $\text{Min}(f)$ as the variance, maximum value, and minimum value of some affine min-tradeoff function $f(q)$.
The min-tradeoff function is an affine function $f(q)$, where for all $i=1,\cdots,N$,~\cite{Metger_EAT2022}
\begin{equation}
    f(q)\leq \min_{\rho\in\Sigma_i(q)} H(X_i|E_i\tilde{E}_{i-1})_{\rho},
\end{equation}
with $\Sigma_i(q)=\{\rho|\rho_{X_iC_iR_iE_i\tilde{E}_{i-1}}=\vars{M}_i(\omega_{R_{i-1}E_{i-1}\tilde{E}_{i-1}}),\rho_{C_i}=q\}$, and $\tilde{E}_{i-1}$ as a purifying system of $R_{i-1}E_{i-1}$. 

\section{Theoretical analysis} \label{sec:theory}

The current methods to analyse sampling techniques~\cite{konig_sampling_2011, dupuis_entanglement_2015} provide bounds which are less than ideal due to their large penalty term, as explained in section \ref{sec:intro}. 
Here, we propose a different method of sampling, which is applicable to a general class of protocols that we term GEAT-analysable protocols. 
By capitalising on the properties of this class of protocols, we demonstrate that the penalty due to sampling is greatly reduced relative to current techniques.
Therefore, for bit strings that are generated by this class of protocols, users can simply switch the randomness extraction step to the sample and hash method proposed here to achieve improved performance.\\

\subsection{Generic GEAT-analysable protocol}

Here, we introduce briefly the class of protocols that are amenable to the sample and hash method, termed GEAT-analysable protocol.
Most critically, the protocol can be proven to be secure using the generalised EAT framework.
Many QKD and QRNG protocols can be analysed with this framework, with Metger et al.~\cite{Metger_QKD2022} providing a set of instructions on how this analysis can be performed on a generic QKD protocol.\\

Consider a generic $N$-round protocol, with honest parties (or a single honest party) and an adversary, 
\begin{enumerate}
    \item \textbf{Data Generation}: The honest parties exchange classical and quantum information for $N$ rounds, which could be attacked by the adversary. 
    At the end of each round of exchange, the honest parties receive a raw bit string $A_i$, with the honest parties publicly exchanging information $I_i$, and with the adversary holding side information $E_i$.
    We assume that $A_i$ is generated independently from any memory (from previous rounds) that the honest parties possesses.
    \item \textbf{Statistical Check}: Honest parties assign each round of the data generation as either a test round ($T_i=1$) or a data generation round ($T_i=0$), with the testing probability $\gamma$. For test rounds, the honest parties announce information $I_i^{test}$, which includes $A_i$, and compute statistics $C_i$. Based on the observed statistics $C^N$, the protocol aborts if they do not fall into the set of events $\Omega$.
    \item \textbf{Additional information exchange}: Additional public information can be exchanged by the honest parties, labelled $Y$, which are not generated in a round-by-round manner (for instance, error correction).
    \item \textbf{Sifting}: Honest users discard rounds which are inconclusive (for instance due to loss or no detection) and part of the test rounds, arriving at a shorter sifted bit string $A'$.
    \item \textbf{Randomness Extraction}: Randomness extraction is performed on the sifted bit string $A'$ to obtain output $K=f_{RE}(A')$, where $f_{RE}$ is a hash function randomly chosen from a family of 2-universal hash functions.
\end{enumerate}
At the end of the protocol, the adversary would have information $E_N'=T^NI^N(I^{test})^NE_N$, along with $Y$ and the hash function $F_{RE}$.\\

After randomness extraction, one ideally wants the hashed output $K$ to be secret from the adversary.
Using the composable security framework~\cite{canetti_universally_2001,hutchison_universal_2005}, the secrecy of bit string $K$ can be defined by
\begin{equation}
    p_{\Omega}\Delta_t(\rho_{KF_{RE}E_N'Y|\Omega},\tau_K\otimes\rho_{F_{RE}E_N'Y|\Omega})\leq\varepsilon,
\end{equation}
where $\Delta_t(\rho,\sigma)$ is the trace distance between states $\rho$ and $\sigma$.
Using the quantum leftover hash lemma~\cite{Tomamichel2011,Tomamichel2017} and chain rule of smooth min-entropy~\cite{Tomamichel2016,Vitanov2013}, proving the security can be reduced to finding a lower bound on the smooth min-entropy $H_{min}^{\varepsilon_{sm}}(A^{'N}|E_N')_{\rho_{A^{'N}C^NR_NE_N'|\Omega}}$. 
Since the protocol can be analysed with GEAT, one expects that the min-entropy and thus security can be computed directly using GEAT, as expressed in the form of Eqn.~\eqref{eqn:GenEATEqn}, providing a secrecy of 
\begin{equation}
    \varepsilon=2\varepsilon_{sm}+\frac{1}{2}\times 2^{-\frac{1}{2}[Nh-v_1\sqrt{N}-v_0-\log_2\abs{\vars{Y}}-l]},
\end{equation}
where $\log_2\abs{\vars{Y}}$ is the length of $Y$.
We note here that tighter bounds on the penalty due to the announcement of additional information $Y$ can be provided, for instance using tighter min-entropy chain rules, but the analysis would be protocol dependent.

\subsection{Sub-block hashing protocol}

In this section, we present our sampled sub-block hashing protocol, which is meant to replace the sifting and randomness extraction step of the generic protocol.
One begins with the raw bit string $A^N = (A_1,\dots,A_N)$ that is supposed to undergo sifting and randomness extraction. 
Let $p_S$ be the sampling probability and $N_S = \frac{1}{p_S}$ be the number of sampled sub-blocks.
Assume, for simplicity, that $p_S$ is chosen such that we have an integer number of sub-blocks, $N_S \in\mathbb{N}$. 

\begin{enumerate}
    \item \textbf{Sampling}: 
    For each round $i\in[1,N]$, the honest parties generate a uniform random variable $V_i$, which can take values $v_i\in [1,N_S]$ (i.e. $V_i$ can take on values between 1 to $N_S$, each with probability $p_S$). 
    If $V_i=j$, we say that bit $A_i$ is ``sampled" into the $j$-th sub-block, which we denote as $A_{S_j}$, where the indices for a set $S_j=\{i:V_i=j\}$.
    The value of $V_i$ is then announced to the rest of the honest parties (if necessary).

    \item \textbf{Sifting}: For each sub-block $S_j$, sifting can be performed to discard rounds which are inconclusive or part of the test rounds, leaving us with sifted sub-blocks $A_{S_j}'$.
    In general, the length of these bit strings are not fixed.
    Therefore, to prevent an excessively long bit string from slowing the Toeplitz hashing, one can choose to abort if the size of any sifted sub-blocks $A_{S_j}'$ exceeds a pre-determined threshold $L_S^{\text{UB}}$.
    In practice, this aborting probability would be very small.    
    
    \item \textbf{Randomness Extraction}: If the protocol does not abort, randomness extraction is performed on all the sub-blocks independently to yield output $K_j=f_{RE,j}(A_{S_j}')$, where $f_{RE,j}$ are random hash functions chosen independently from a family of 2-universal hash functions.
    For simplicity, we take the length of output $K_j$ to be $\tilde{l}$ for every $j$. 
    
    \item \textbf{Concatenation}: The final output, denoted by $K=(K_1,\cdots,K_{N_S})$, is obtained by concatenating all the individual outputs of randomness extraction on each sub-block.
\end{enumerate}
When the sifting and randomness extraction is replaced with sampled sub-block hashing, $V_i$ is additionally announced, along with the use of multiple hash function, $f_{RE,1}\cdots f_{RE,N_S}$, and would be accessible to the adversary.

\subsection{Security of sub-block hashing}

Replacing the sifting and randomness extraction with sampled sub-block hashing, one remains concerned about the security of the bit string $K$, now formed from a concatenation of sub-strings.
The security of the protocol is closely tied to the security of each sub-string $K_j$, which one could consider as being generated from a sub-protocol.
The security of this sub-protocol, as we claim below, can be easily computed based on the security analysis of the original GEAT-analysable protocol.
We note while the security claim provided here deals with the ease of computing the secrecy of the protocol when adapted with sub-block hashing, it makes no assertion on whether this would cause an increase in performance.
\begin{claim}
Replacing the sifting and randomness extraction with the sampled sub-block hashing steps, the security of the protocol 
\begin{equation}
    p_{\Omega}\Delta_t(\rho_{KV^NE_N'F_{RE}^{N_S}Y|\Omega},\tau_K\otimes\rho_{V^NE_N'F_{RE}^{N_S}Y|\Omega})\leq \varepsilon,
\end{equation}
can be computed from the security of the sub-protocols, $\varepsilon=N_S\varepsilon'$.
The sub-protocol security $\varepsilon'$ can be computed based on its associated smooth min-entropy, with simple replacement of the min-tradeoff function $f(q)\rightarrow \frac{1}{N_S}f(q)$ and sub-string size $l\rightarrow \tilde{l}=\frac{l}{N_S}$, i.e.
\begin{equation}
    \varepsilon'=2\varepsilon_{sm}+\frac{1}{2}\times 2^{-\frac{1}{2}[Nh'-v_1'\sqrt{N}-v_0'-\log_2\abs{\vars{Y}}-\tilde{l}]},
\end{equation}
where $h'$, $v_0'$ and $v_1'$ are computed with the min-tradeoff function replacement.
\end{claim} 

Here, we provide a brief idea of the security analysis demonstrating the claim, with the full proof provided in Appendix~\ref{App:DetailedSecProof}.
For each sub-protocol, the associated smooth min-entropy is of the form $H_{min}^{\varepsilon_{sm}}(A_{S_j}'|A_{S_{j+1}}'\cdots A_{S_{N_S}}'V^NE_N')$, which accounts for correlations between the sampled raw bit strings.
Since the original protocol is GEAT-analysable, one is able to write down EAT channels $\vars{M}_i$ and apply the EAT theorem by defining some min-tradeoff function $f(q)$.
To construct the min-tradeoff function for the sub-protocol, one can adapt the EAT channel to output $A_{S_j,i}'$ (note $A_{S_j,i}'=\perp$ is inserted if the $i$-th round is discarded/not included in $A_{S_j}'$) while forwarding the information of $A_{S_{j+1},i}'\cdots A_{S_{N_S},i}'$ and $V_i$ to Eve.
Since only $p_S$ fraction of the bits $A_i'$ are output as $A_{S_j,i}'$, the entropy and thus the min-tradeoff function gains a factor of $p_S=\frac{1}{N_S}$, giving the modification in the claim.\\

We note again that the penalty due to announcement of additional information exchanged, $Y$, could be tighter bounded, depending on the protocol. 
For instance, if $Y$ is the syndrome sent for error correction, adapting the error correction to apply to each sub-block could reduce the penalty, since each sub-block $A_{S_j}'$ would only face penalty due to error correction for that sub-block itself, $Y_j$.
This could in principle reduce the error correction rate by a factor of $N_S$ on each sub-block.
We also note that for protocols with additional parameter checks, such as the key verification step in QKD to ensure correctness, the additional conditioning on the event that these checks pass, $\Omega_{KV}$, can be removed by a simplification of the smooth min-entropy: $H^{\varepsilon_{sm}}_{min}(A^N|E_N)_{(\land\Omega_{KV})|\Omega}\geq H^{\varepsilon_{sm}}_{min}(A^N|E_N)_{|\Omega}$ for $\varepsilon_{sm} \leq \sqrt{\Pr[\Omega_{KV}]}$~\cite{Tomamichel2017}. 

\subsection{Excessive sub-block size}

The sampled sub-blocks would in general be shorter in length (by a factor of $N_S$) and thus enjoy a speed-up during the hashing process.
However, due to the nature of bit-wise sampling, there remains a small probability that the length of one sub-block could be excessively long.
To properly account for these events in our analysis, we set a block size limit, $L_S^{\text{UB}}$, where if any sub-block exceeds, the protocol would abort.
As such, one can be sure that if the protocol succeeds, the speed-up can be guaranteed by the fact that the hashing process has an input block size of at most $L_S^{\text{UB}}$.\\

Aborting when the bit strings are excessively long incurs a small penalty to the secrecy, $\varepsilon_{\Omega'}$, the probability that these rare events occur, i.e., $\Pr[\cup_j\{|A_{S_j}'|>L_S^{\text{UB}}\}]\leq \varepsilon_{\Omega'}$.
One could in principle select the block size limit with any concentration bound, and here we demonstrate with a concentration bound on binomial distribution~\cite{Zubkov2013}, that one can choose the block size limit such that
\begin{equation}
    \frac{\varepsilon_{\Omega'}}{m}=1-\Phi\left[\sqrt{2NH\left(\frac{\lfloor L_S^{\text{UB}}\rfloor}{N},p_{sift}\right)}\right],
\end{equation}
where $p_{sift}$ is the probability that a protocol would output a bit from $A_i$ into $A_{S_j}'$ (has to be conclusive, not be a test round and sampled into block $S_j$), $\Phi(x)$ is the normal cumulative distribution function and $H(x,p)=x\ln(x/p)+(1-x)\ln[(1-x)/(1-p)]$.
Including this aborting chance would result in an additional $\varepsilon_{\Omega'}$ penalty to the security condition, since the trace distance between the state including and excluding this aborting part is bounded by $\varepsilon_{\Omega'}$.

\section{Results} \label{sec:results}

We demonstrate our method by employing it on a simulated standard BBM92~\cite{bennett_quantum_1992} QKD protocol with $\varepsilon_{\Omega'} = 1\times 10^{-8}$, $e_{\text{ph}}=0.82\%$, and $e_{\text{bit}}=5.8\%$. The error rates are chosen based on previously reported experimental results~\cite{erven_entangled_2009}. We then use GEAT to optimise the testing probability $p_{\sX}$ under different simulation parameters. We note that the different $p_{\sX}$ values can be easily achievable experimentally using passive beam splitters with fixed splitting ratio, or using optical switches for active basis choice.

\begin{figure}
    \centering
    \includegraphics[width=\columnwidth]{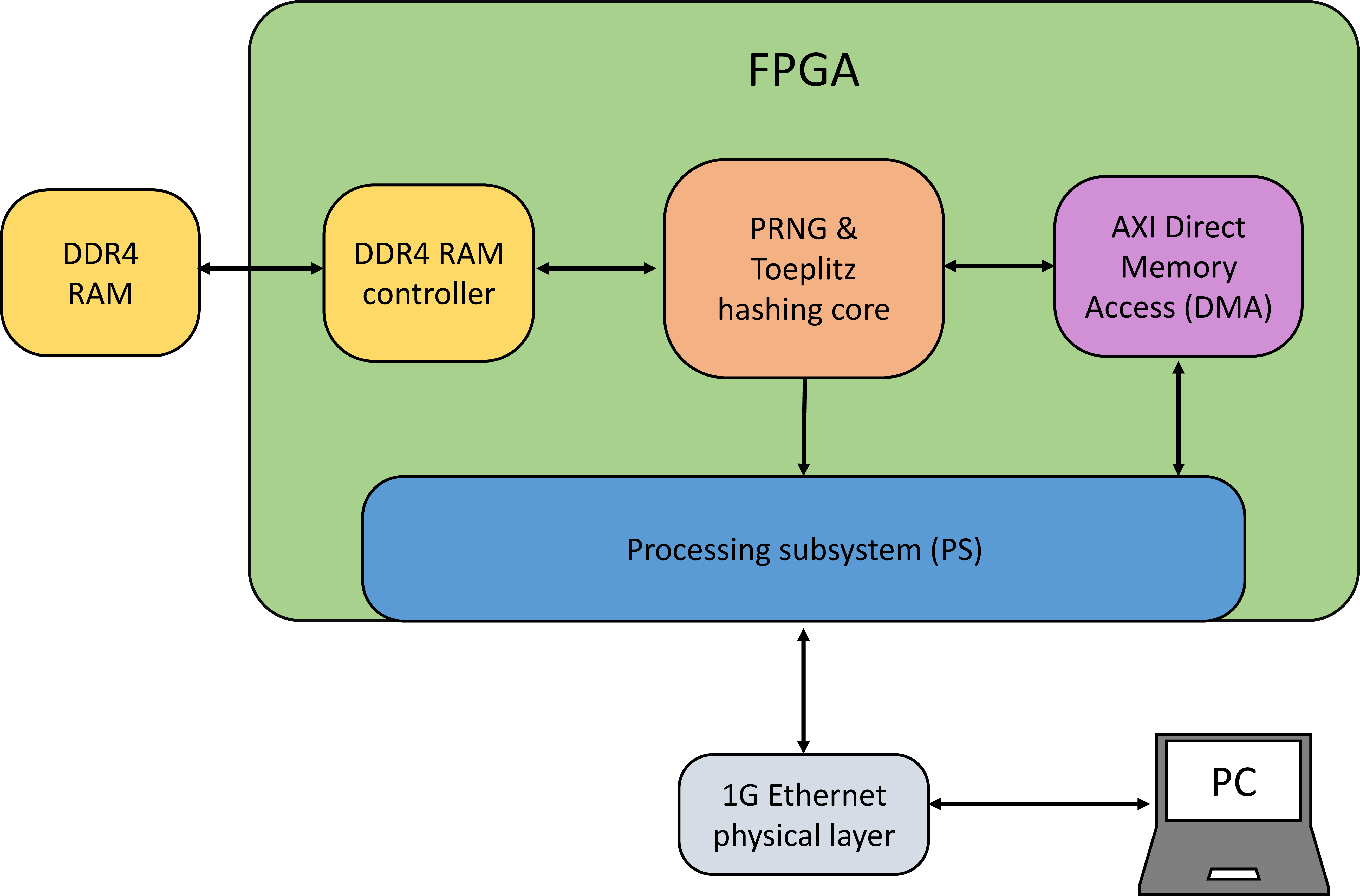}
    \caption{Schematic of Toeplitz hashing on FPGA using our proposed method. To perform the test without our method, the PRNG \& Toeplitz hashing module is switched out to another module that performs only Toeplitz hashing.}
    \label{fig:schematic}
\end{figure}

We first design a module on FPGA to carry out Toeplitz hashing.  
The schematic of the hardware module on FPGA is shown in Fig. \ref{fig:schematic}. We explain the process for sampling of the first sub-block, and note that the implementation for the rest of the sub-blocks is largely similar, with some minor changes. The hashing module receives one bit of input data every clock cycle and runs a PRNG algorithm. If the output of the PRNG is within a certain range, the module stores the bit in external DDR memory (i.e., the bit is sampled). Once the sampling is done, the module receives one bit of seed data from the processing subsystem (PS) and one bit of raw data from the DDR RAM every computation cycle. The seed bits are accumulated as a seed string and the string is used for matrix-vector multiplication. At the end of the computation cycle, a shift register is used to shift the accumulated seed string by one position. The string will then be ready to receive the new seed bit that is incoming in the next computation cycle. 

The module also splits the large Toeplitz matrix to a smaller matrix of size $m' \times n'$ and performs the matrix-vector multiplication iteratively. We use a pipeline implementation, where the interval between every computation cycle is set to one clock cycle. With this, the estimate for the number of clock cycles required to perform hashing on one sub-block is approximately $L_S^{\text{UB}} + \tilde{l} + (L_S^{\text{UB}} + m') \times \frac{\tilde{l}}{m'}$. For our implementation, we set $m'=2000 \text{ and } n'=1$.

\subsection{Simulation with fixed testing probability} \label{sec:fixedPx}

\begin{figure*}[!th]
    \centering
    \subfloat[\label{fig:KeyRatePerSignal}]{\includegraphics[width = \columnwidth]{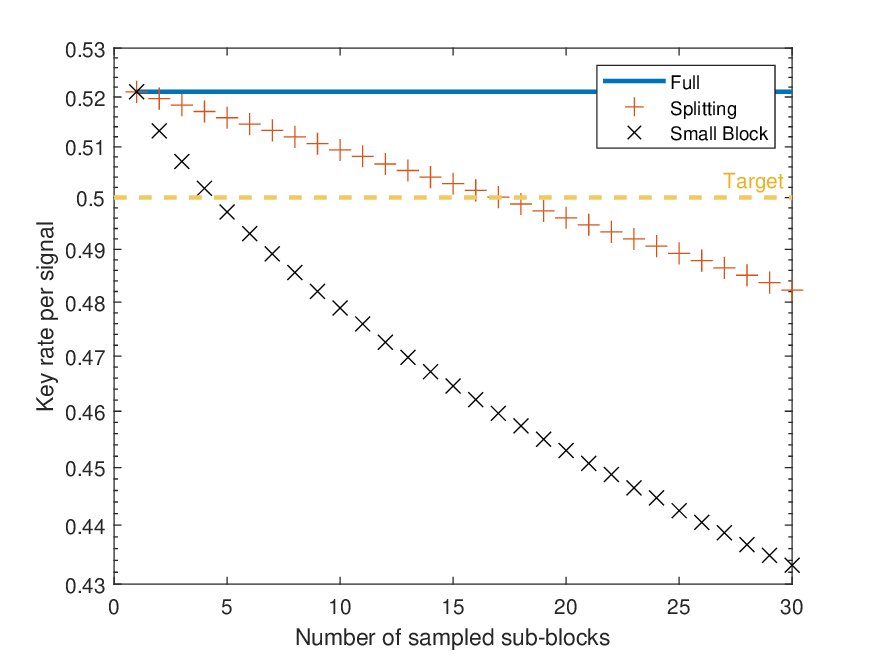}
    }
    \subfloat[\label{fig:KeyRatePerTime}]{\includegraphics[width = \columnwidth] {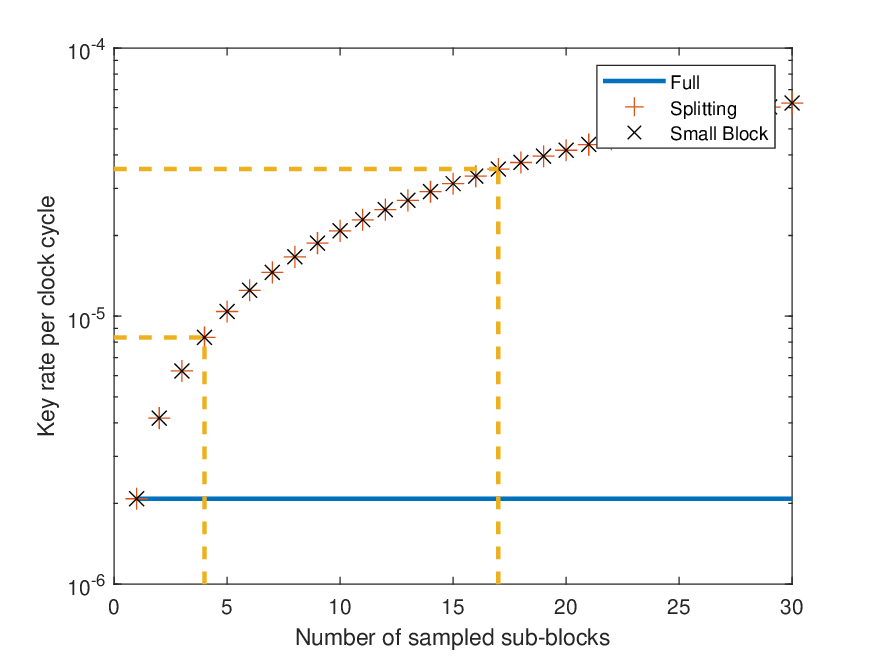}
    } \\
    \caption{Simulated key rate results for a standard BBM92 QKD protocol, where $N=1\times 10^9$, $e_{\text{ph}}=0.82\%$, $e_{\text{bit}}=5.8\%$, $p_{\sX}=0.02$, and $\varepsilon_{sec}=1\times 10^{-6}$. Full refers to hashing the sifted key directly, splitting refers to using our sampled sub-block hashing method, and small block refers to running the QKD protocol for $(N/N_S)$ rounds for $N_S$ times, and performing parameter estimation and hashing for each of the $N_S$ blocks separately. (a) Plot of key rate per signal sent against the number of sampled sub-blocks. The dashed line represents a target key rate per signal of 0.5. (b) Plot of key rate per unit time against the number of sampled sub-blocks. The dashed lines represent the optimal $N_S$ value, and its corresponding throughput, that achieves the target key rate per signal for the splitting $(N_S=17)$ and the small block $(N_S=4)$ method.
    }
    \label{fig:SimulatedResults}
\end{figure*}

Using the key length presented in Appendix \ref{App:BB84Sec} and the timing information, we simulate the key rate for three scenarios: (1) hashing the sifted key directly, (2) using sampled sub-block hashing, and (3) running the QKD protocol for a smaller number of rounds $(N/N_S)$ for $N_S$ times, totalling $N$ rounds, and performing parameter estimation and hashing for each of the $N_S$ blocks separately. The main result of this paper would be the comparison between scenarios one and two. However, as the last scenario is largely similar to our method, we will make additional comparisons between scenarios two and three throughout the paper as well.

To illustrate the effects of the sampled sub-block hashing method, we consider a BBM92 protocol with $N=10^9$ rounds, and a secrecy parameter of $\varepsilon_{sec} = 1\times 10^{-6}$. The optimised $p_{\sX}$ value for direct hashing is 0.0176, which we round to 0.02. This method of choosing $p_{\sX}$ corresponds to the scenario where an optimised setup has already been achieved, and one would like to modify the post-processing step to improve the throughput. From Fig. \ref{fig:SimulatedResults}, it is clear that using our method, there is a small loss in key rate per signal sent, but we obtain a linear speed-up. The loss in key rate from sampling, as shown in Fig. \ref{fig:KeyRatePerSignal}, is expected, and is a result of the penalty from sampling. Interestingly, despite the loss, there is an advantage over starting with a smaller block. This may be due to the fact that our method uses the overall statistics of $N$ rounds to estimate the min-entropy of each sub-protocol, whereas the small block scenario only uses each protocol's statistics of $N/N_S$ rounds. In addition, as the time required scales more favourably for lower input size (based on the clock cycle equation we obtained), a shorter sub-block would improve generation rate despite a loss in key length, as seen in Fig. \ref{fig:KeyRatePerTime}. 

Now, suppose that the protocol is run with the parameters presented in Fig. \ref{fig:SimulatedResults}, and the user requires a key of length $5 \times 10^8$ bits. This corresponds to a value of 0.5 for the key rate per signal, which is indicated by the dashed line in Fig. \ref{fig:KeyRatePerSignal}. In this scenario, $N_S=4$ is the optimal value for the small block implementation. On the other hand, if our method is utilised, then the optimal value would now be $N_S=17$. From the plot shown in Fig. \ref{fig:KeyRatePerTime}, according to the dashed lines, both methods would result in an improvement in throughput compared to direct hashing. However, our method has a throughput that is higher than the small block implementation by more than a factor of 4. This shows the advantage of our method over the small block implementation. 

\begin{figure*}[t]
    \centering
    \subfloat[\label{fig:Secrecy}]{\includegraphics[width = \columnwidth]{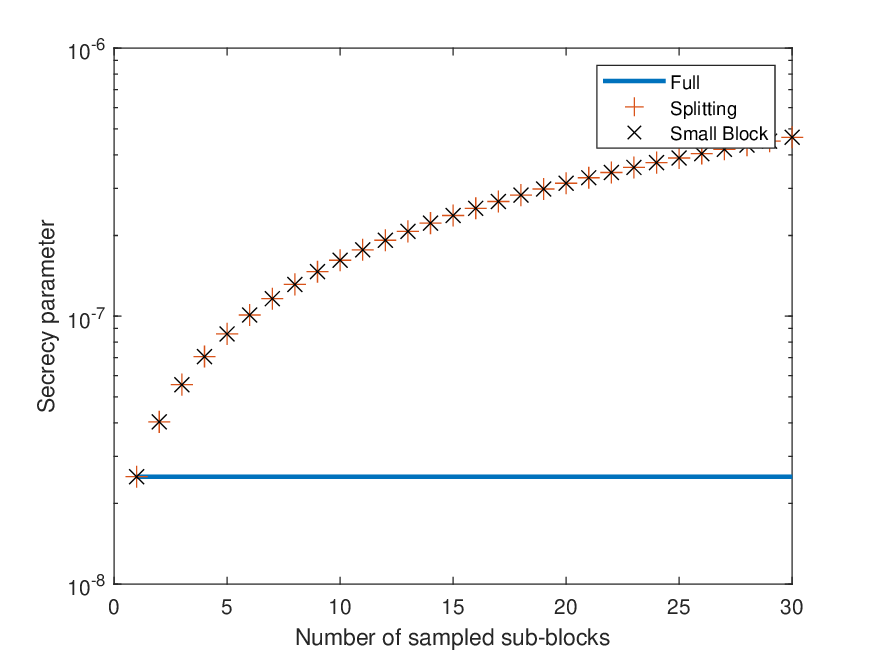}
    }
    \subfloat[\label{fig:SecrecySpeed}]{\includegraphics[width = \columnwidth] {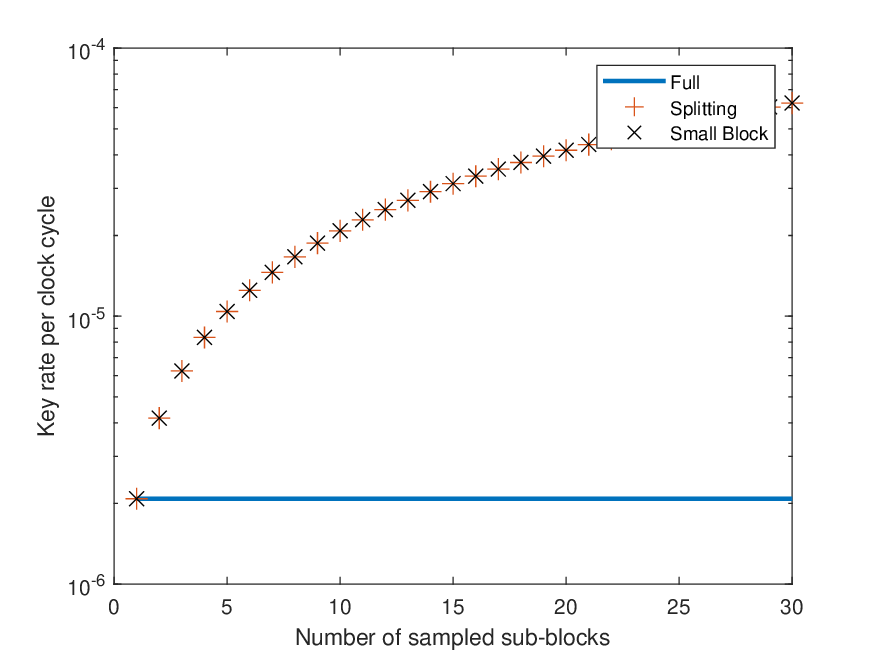}
    } \\
    \caption{Simulated results for a standard BBM92 QKD protocol, where $N=1\times 10^9$, $l=4.3\times 10^8$, $e_{\text{ph}}=0.82\%$, $e_{\text{bit}}=5.8\%$, and $p_{\sX}=0.02$. (a) Plot of optimised secrecy parameter against the number of sampled sub-blocks. (b) Plot of key rate per unit time against the number of sampled sub-blocks.}
    \label{fig:OptimiseSecrecy}
\end{figure*}

To study the behaviour of the secrecy parameter $\varepsilon_{sec}$, we fix the output length to $l=4.3\times 10^8$ bits, $N=10^9$, and $p_{\sX}=0.02$. We optimise the secrecy parameter for all three scenarios under different values of $N_S$ and plot the results in Fig. \ref{fig:OptimiseSecrecy}. From the results, we can see that the sampled sub-block hashing method incurs a small linear loss in terms of secrecy, but allows for a linear improvement in terms of throughput.

\begin{figure}[t]
    \centering
    \includegraphics[width= \columnwidth]{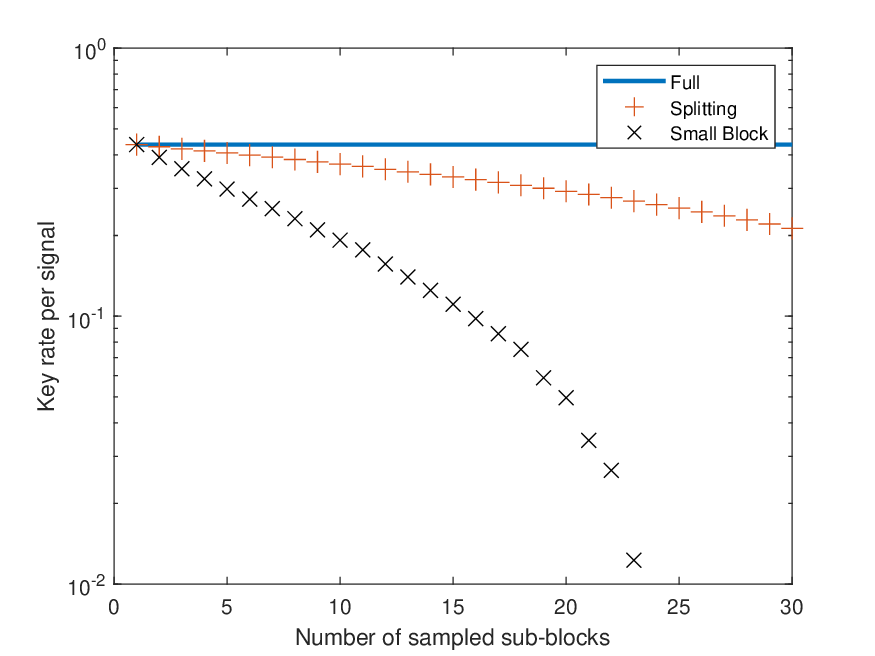}
    \caption{Plot of key rate per signal sent against the number of sampled sub-blocks for a standard BBM92 QKD protocol, where $N=3\times 10^7$, $e_{\text{ph}}=0.82\%$, $e_{\text{bit}}=5.8\%$, $p_{\sX}=0.02$, and $\varepsilon_{sec}=1\times 10^{-6}$.}
    \label{fig:SmallRoundKeyRate}
\end{figure}

We observe that our method is able to achieve positive key rates in parameter regimes that are previously not possible. As an example, Fig. \ref{fig:SmallRoundKeyRate} contains the calculated key rate per signal result when $N=3\times 10^7$, $p_{\sX}=0.02$, and $\varepsilon_{sec}=10^{-6}$. Under the small block implementation, hashing has to be performed on a minimum block size of approximately $1.3\times 10^6$ bits to obtain a positive key rate. Using our method, we are able to perform hashing on an even smaller block size of $1\times 10^6$ bits and obtain a key rate per signal that is higher by more than an order of magnitude.

\subsection{Simulation with optimised testing probability}

\begin{figure*}[t]
    \centering
    \subfloat[\label{fig:KeyRateOptPX1e9}]{\includegraphics[width = \columnwidth]{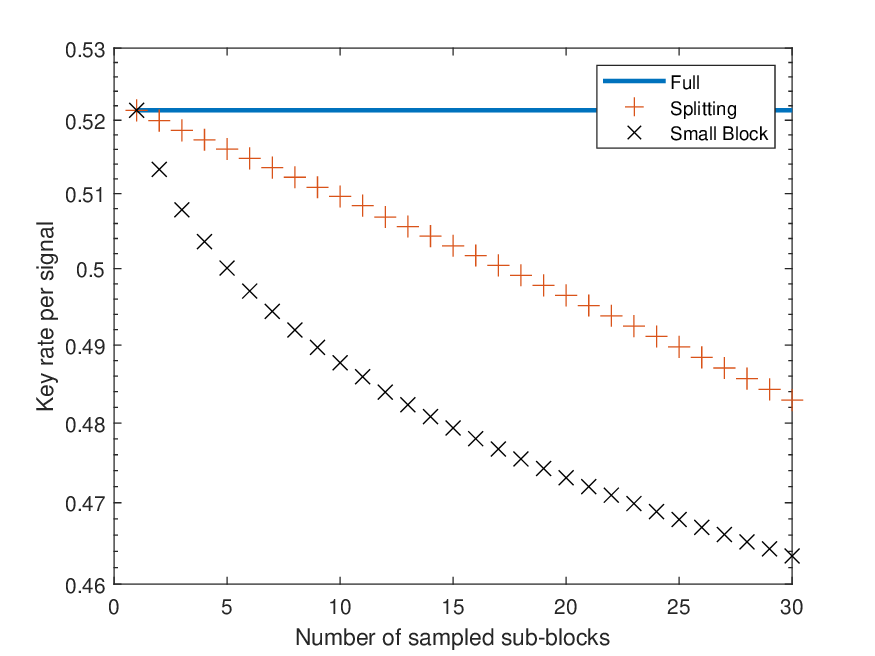}
    }
    \subfloat[\label{fig:SpeedOptPX1e9}]{\includegraphics[width = \columnwidth]{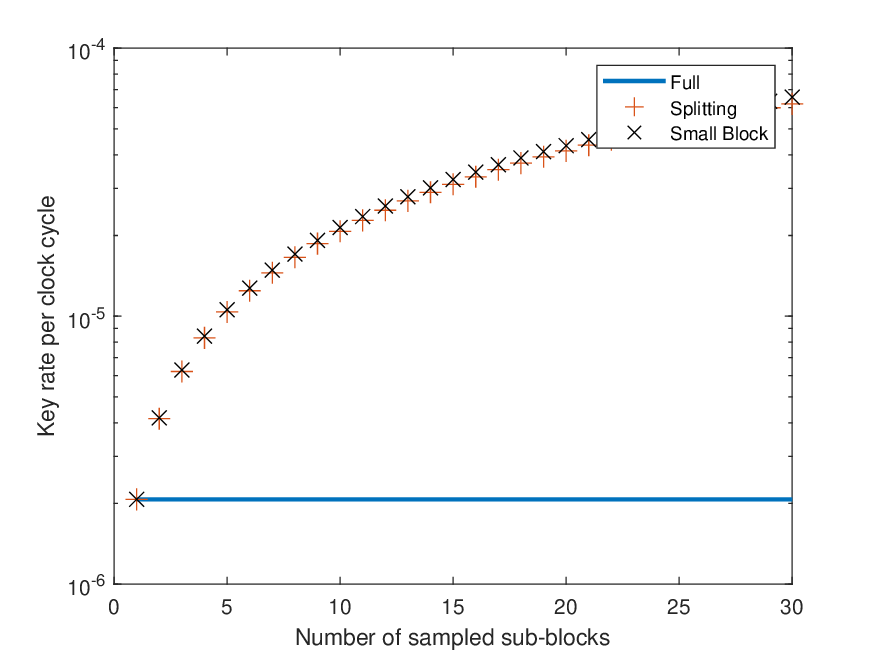}
    }\\
    \subfloat[\label{fig:OptPX1e9}]{\includegraphics[width = \columnwidth]{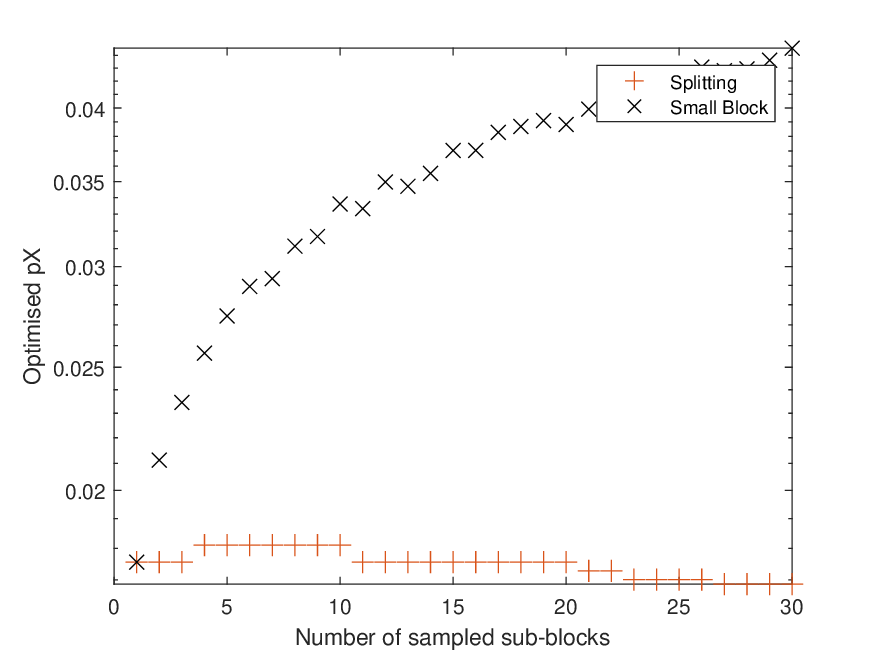}
    }\\
    \caption{Simulated results for a standard BBM92 QKD protocol, where $N=1\times 10^9$, $e_{\text{ph}}=0.82\%$, $e_{\text{bit}}=5.8\%$, and  $\varepsilon_{sec}=1\times 10^{-6}$. The sampling probability $p_{\sX}$ is left as a free parameter to be optimised. (a) Plot of key rate per signal sent against the number of sampled sub-blocks. (b) Plot of key rate per unit time against the number of sampled sub-blocks. (c) Plot of optimal $p_{\sX}$ against the number of sampled sub-blocks.}
    \label{fig:SimulationOptimisePX1e9}
\end{figure*}

Next, we consider the case where the testing probability $p_{\sX}$ is not fixed (e.g., the setup is still in the design optimisation stage, or $p_{\sX}$ can be easily changed with an active beam splitter). We again present the simulated key rate results for the same three scenarios mentioned in section \ref{sec:fixedPx}. For this case, we optimise $p_{\sX}$ over all values of $N_S$, for all three scenarios. The simulated results for $N=1 \times 10^9$ rounds and $\varepsilon_{sec}=1 \times 10^{-6}$ are shown in Fig. \ref{fig:SimulationOptimisePX1e9}. Comparing Figs. \ref{fig:KeyRatePerSignal} and \ref{fig:KeyRateOptPX1e9}, our method achieves similar key rate per signal for both cases. On the other hand, for the case of small block implementation, optimising the value of $p_{\sX}$ provides an 
improvement to the key rate per signal. This is because the optimal $p_{\sX}$ increases with respect to $N_S$, as shown in Fig. \ref{fig:OptPX1e9}, which allows for a larger parameter estimation set. Hence, there is less penalty from statistical fluctuations. However, our method is still able to achieve a higher key rate than the small block implementation in this parameter regime. 

\begin{figure*}[t]
    \centering
    \subfloat[\label{fig:KeyRateOptPX3e7}]{\includegraphics[width = \columnwidth]{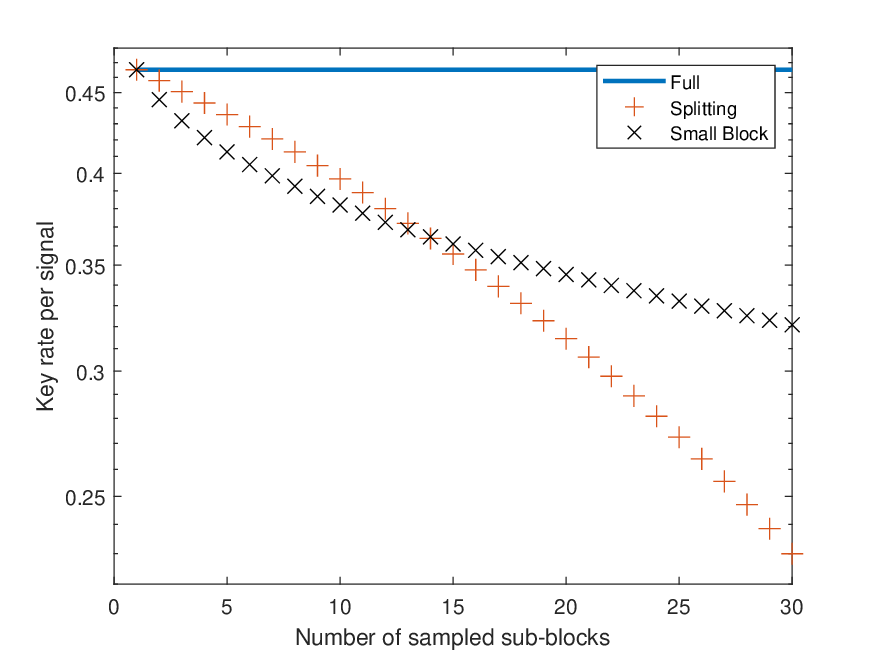}
    }
    \subfloat[\label{fig:OptPX3e7}]{\includegraphics[width = \columnwidth]{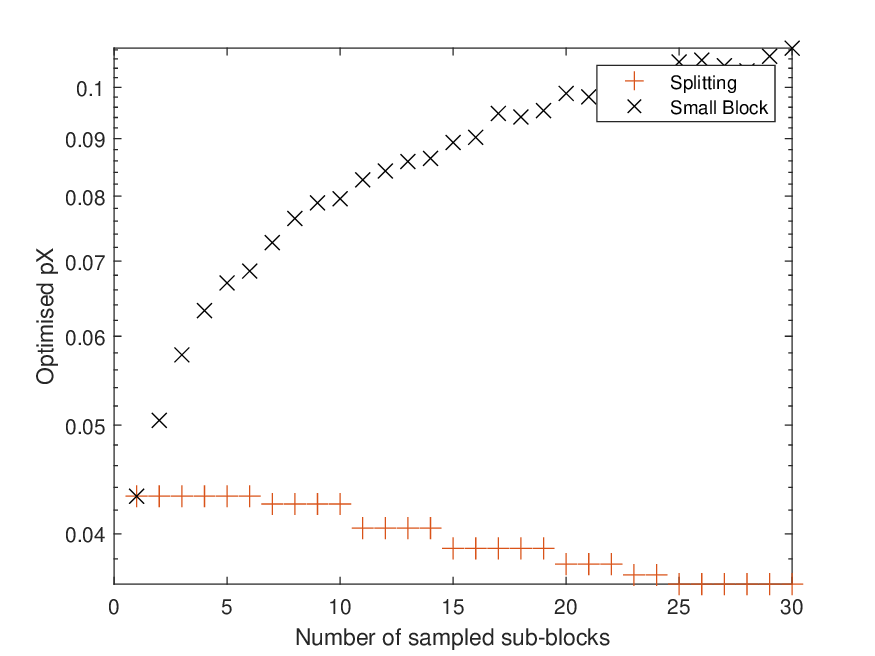}
    }\\
    \caption{Simulated results for a standard BBM92 QKD protocol, where $N=3\times 10^7$, $e_{\text{ph}}=0.82\%$, $e_{\text{bit}}=5.8\%$, and  $\varepsilon_{sec}=1\times 10^{-6}$. The sampling probability $p_{\sX}$ is left as a free parameter to be optimised. (a) Plot of key rate per signal sent against the number of sampled sub-blocks. (b) Plot of optimal $p_{\sX}$ against the number of sampled sub-blocks.}
    \label{fig:SimulationOptimisePX3e7}
\end{figure*}

Similarly, we optimised the $p_{\sX}$ values for $N=3 \times 10^7$ and $\varepsilon_{sec}=1 \times 10^{-6}$, and plot the simulated key rate results in Fig. \ref{fig:SimulationOptimisePX3e7}. We see that in this case, the small block implementation will outperform our method when the value of $N_S$ is greater than 13, as shown in Fig. \ref{fig:KeyRateOptPX3e7}. Comparing Figs. \ref{fig:OptPX1e9} and \ref{fig:OptPX3e7}, the optimal $p_{\sX}$ increases with $N_S$ for the small block implementation for both cases. As explained, this is to allow for a larger parameter estimation set, and therefore reduce the penalty from statistical fluctuations. For our method, since the overall string is utilised for parameter estimation, the effect of statistical fluctuation remains relatively unchanged with increasing number of sub-blocks, resulting in similar optimal $p_{\sX}$ values. The decrease in key rate is thus primarily due to penalty from the sampling method, which does not appear to be sensitive to $p_{\sX}$.

\subsection{Hardware implementation}

We implemented large input Toeplitz hashing on FPGA with and without our method, using simulated datasets. For our method, although the sampling procedure is executed in full, we perform privacy amplification on only the first three sub-blocks independently, before concatenating the hashed outputs together as a proof of concept demonstration. 

\begin{table}[t]
\resizebox{\columnwidth}{!}{%
\begin{tabular}{cccc}
\toprule
\multirow{2}{*}{\begin{tabular}[c]{@{}c@{}}Input size\\ (Mbits)\end{tabular}} & \multirow{2}{*}{\begin{tabular}[c]{@{}c@{}}Output size\\ (Mbits)\end{tabular}} & \multicolumn{2}{c}{Speed (bits/s)} \\ \cline{3-4} &  & Traditional & Our method \\ \hline
96.040    & 6.054   & 498.60    & 9,899.80        \\
192.08    & 13.302  & 249.31    & 4,972.47        \\
288.12    & 20.736  & 166.24    & 3,318.63        \\
960.40    & 74.406  & -         & 997.05          \\
1,920.00  & 152.56  & -         & 498.62          \\ \bottomrule
\end{tabular}}
\caption{Tabulated results for hardware implementation of Topelitz hashing with and without our method.}
\label{tab:FPGAResults}
\end{table}

We choose $p_S=0.05$, because at that point, the loss in extractable key length is small, but the increase in execution speed is large. We used a Xilinx ZCU111 evaluation board as our implementation platform and tabulated our results in Table~\ref{tab:FPGAResults}. As expected, when the input size of privacy amplification increases, the speed of the privacy amplification decreases linearly. This is because to generate even a single bit of output, a larger number of bits needs to be processed first. When comparing between direct hashing (traditional) and our method, it is clear that privacy amplification with our method has an execution speed that is faster by almost 20 times. This shows that our method allows one to satisfy the finite-size effect without too heavy a penalty on speed. In particular, for the cases where the input size is 960.40 Mbits and 1920 Mbits, Toeplitz hashing without our method would have taken approximately 413 hours and 1695 hours respectively, but our method reduced the timing to 20.73 hours and 84.99 hours, therefore greatly improving the practicality of large input hashing. 

\begin{figure*}[t]
    \centering
    \subfloat[\label{fig:PValue}]{\includegraphics[width = \columnwidth]{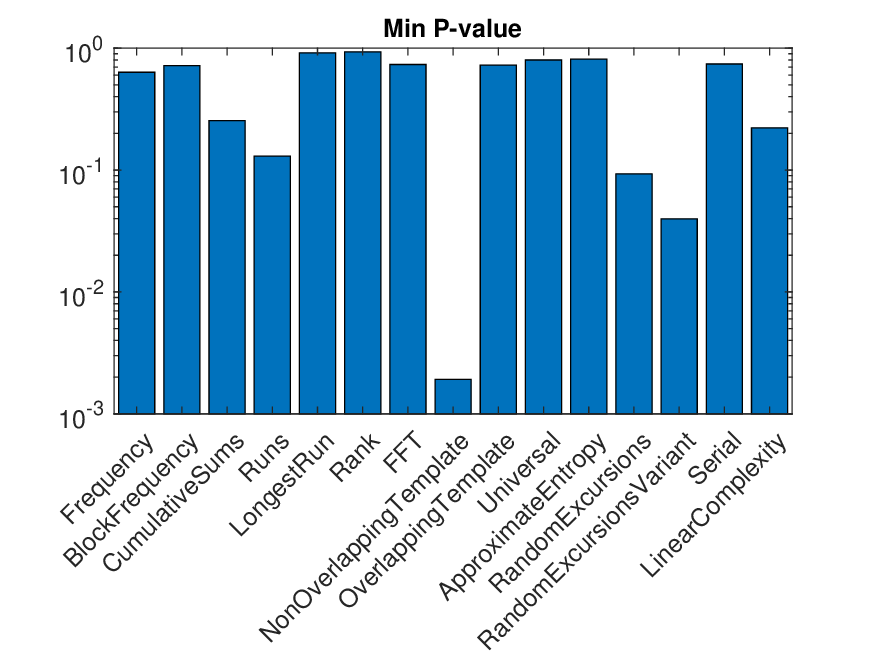}
    }
    \subfloat[\label{fig:Proportion}]{\includegraphics[width = \columnwidth]{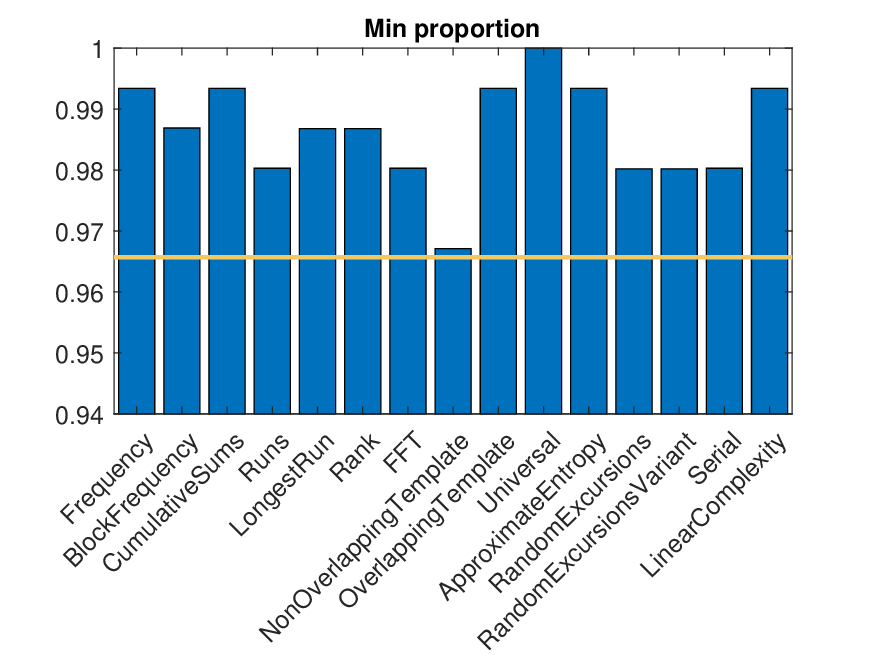}
    } 
    \caption{Results obtained from the NIST 800-22 statistical test suite for the concatenated output of 152.56 Mbit. For each test in the test suite, we plot (a) the P-value, and (b) the proportion of samples that passed. If there were multiple runs of the same test, the minimum value is taken.}
    \label{fig:NistResults}
\end{figure*}

For the final implementation, the concatenated output of 152.56 Mbit is fed into the NIST 800-22 statistical test suite~\cite{rukhin_nist_2010} to test for uniformity. In Fig. \ref{fig:PValue}, we show the P-value results of each individual test in the test suite. A P-value greater than $10^{-4}$ indicates that the sequence under test passes that particular test for uniformity. Additionally, in Fig. \ref{fig:Proportion}, we show the proportion result of each individual test. The proportion result shows the proportion of samples from the sequence under test that passes that particular test. A proportion value greater than 0.9657 \footnote{For random excursion variant test, the value is 0.9603}, illustrated by the horizontal orange line in Fig. \ref{fig:Proportion}, indicates that the sequence under test passes that particular test for uniformity. The combined results show that our concatenated output passes all the tests in the test suite. In addition, as the security is guaranteed by our theoretical model, we can thus confidently certify that the output is uniform and secret, even from the adversary.

Finally, in Table \ref{tab:utilisation}, we present the resource utilisation results from the implementation of our method on the ZCU111 evaluation kit. The utilisation of look-up table (LUT), LUTRAM, flip-flop (FF), block RAM (BRAM), and digital signal-processing (DSP) blocks are low. The low utilisation percentage also indicates that the entire project can fit on platforms which are smaller and have lower cost.

\begin{table}[t]
    \centering
    \begin{tabular}{c c c}
    \toprule
    Resource & Utilisation & Utilisation (\%) \\
    \midrule
    LUT & 60,950 & 14.33 \\
    LUTRAM & 6,184 & 2.90 \\
    FF & 78,685 & 9.25 \\
    BRAM & 32.5 & 3.01 \\
    DSP & 39 & 0.91 \\
    \bottomrule
    \end{tabular}
    \caption{Tabulated resource utilisation for hardware implementation of Toeplitz hashing and concatenation for sampling of 3 sub-bocks.}
    \label{tab:utilisation}
\end{table}

\section{Discussion and conclusion} \label{sec:conclusion}

In this paper, we propose a sampled sub-block hashing method for large block size randomness extraction with low resource utilisation level. Our simulation and experimental results, based on the BBM92 QKD protocol, demonstrate an improvement in key rate per unit time (throughput) compared to both the direct hashing approach and the small block implementation approach, within certain parameter regime of $p_{\sX}$ and $N$. We attribute this improvement to the parameter estimation advantage of our scheme -- parameter estimation is conducted based on the entire $N$ rounds rather than individual small blocks, thereby introducing a smaller penalty term compared to that introduced by sampling. As a result, we conjecture that our method would exhibit a more distinct advantage for protocols requiring more extensive parameter estimation. We also note that although this work only considers direct Toeplitz hashing on FPGA, it is highly plausible that our method would also improve the throughput when applied to other implementations on other platforms (e.g., Toeplitz hashing accelerated by FFT on CPU). We leave these explorations for future work. 

In summary, the proposed randomness extraction method has a straightforward operation and can be applied to a generic class of quantum cryptographic protocols. Most importantly, it offers a remarkable order-of-magnitude improvement in throughput while utilising system resources efficiently. Consequently, our method presents a highly promising solution for achieving high-speed randomness extraction under stringent resource constraints. This feature holds tremendous importance for practical and ultra-high-speed realisations of quantum cryptographic protocols, particularly in the context of on-chip deployments.

\section*{Acknowledgements} 

We acknowledge funding support from the National Research Foundation of Singapore (NRF) Fellowship grant (NRFF11-2019-0001) and NRF Quantum Engineering Programme 1.0 grant (QEP-P2).

\section*{Author contributions}

C. L. designed the research. C. W. and H. J. N. designed and implemented the method on hardware. W. Y. K. and I. W. P. developed the security proof. W. Y. K. did the numerical simulation. H. J. N. and W. Y. K. wrote the manuscript, with contributions from all authors.

\section*{Data availability} 

All the data that support the findings of this study are available in the main text. Source data are available from the corresponding author on reasonable request.

\section*{Code availability}

The codes used for simulation are available from the corresponding author on reasonable request.

\bibliography{biblio}

\appendix

\section{Generalised EAT}
\label{app:GEAT}

The generalised EAT can be stated as
\begin{thm}[Generalised EAT~\cite{Metger_EAT2022}]
    Consider EAT channels $\{\vars{M}_i\}_i$ satisfying the non-signalling and projective reconstructability properties with classical $C_i$ with alphabet $\vars{C}$. Let $\varepsilon\in(0,1)$, $\alpha\in(1,\frac{3}{2})$, $\Omega\subset \vars{C}^N$, and $f$ be an affine min-tradeoff function, then
    \begin{multline}
        H_{min}^{\varepsilon}(X^N|E_N)_{\rho_{X^NC^NR_NE_N|\Omega}}\geq Nh-N\frac{(\alpha-1)\ln 2}{2(2-\alpha)}V^2\\
        -\frac{g(\varepsilon)+\alpha\log_2(\frac{1}{p_{\Omega}})}{\alpha-1}-N(\frac{\alpha-1}{2-\alpha})^2K'(\alpha),
    \end{multline}
    where 
    \begin{gather*}
        h=\min_{c^N\in\Omega}f(\text{\emph{freq}}(c^N))\\
        g(\varepsilon)=-\log_2(1-\sqrt{1-\varepsilon^2})\\
        V=\log_2(2d_X^2+1)+\sqrt{2+\Var(f)}\\
        K'(\alpha)=\frac{(2-\alpha)^3}{6(3-2\alpha)^3\ln 2}2^{\frac{\alpha-1}{2-\alpha}(2\log_2d_X+\text{\emph{Max}}(f)-\text{\emph{Min}}_{\Sigma}(f))}\\
        \times\ln^3(2^{2\log_2 d_X+\text{\emph{Max}}(f)-\text{\emph{Min}}_{\Sigma}(f)}+e^2),
    \end{gather*}
    $p_{\Omega}$ is the probability that the event $\Omega$ occurs and $d_X$ is the dimension of $X_i$.
\end{thm}
One could perform simplification of the generalised EAT~\cite{Metger_EAT2022} to obtain the form in the main text, where the additional terms are
\begin{gather*}
    \xi=\frac{2\ln 2}{1+2\ln 2}\\
    \beta(\varepsilon)=\frac{(2-\xi)\xi^2\log_2(1/p_{\Omega})+\xi^2g(\varepsilon)}{3(\ln 2)^2(2\xi-1)^3}\\
    \gamma(\varepsilon)=\sqrt{\frac{2\ln 2}{\xi}[g(\varepsilon)+(2-\xi)\log_2(1/p_{\Omega})]}.
\end{gather*}

\section{Detailed Security Analysis of Sub-block Hashing}
\label{App:DetailedSecProof}

In this section, a detailed analysis of the security of sub-block hashing is provided.
As mentioned in the main text, the security condition of concern is
\begin{equation}
    p_{\Omega}\Delta_t(\rho_{KV^NE_N'F_{RE}^{N_S}Y|\Omega},\tau_K\otimes\rho_{V^NE_N'F_{RE}^{N_S}Y|\Omega})\leq \varepsilon',
\end{equation}
where one can guarantee the bit string $K$ is secret from the adversary, who has access to $V^NE_N'F_{RE}^{N_S}Y$.
By introducing an intermediate state of the form $\tau_{K_1\cdots K_j}\otimes\rho_{K_{j+1}\cdots K_{N_S}V^NE_N'F_{RE}^{N_S}Y|\Omega}$, one can apply the triangle inequality to arrive at
\begin{multline}
    p_{\Omega}\Delta_t(\rho_{KV^NE_N'F_{RE}^{N_S}Y|\Omega},\tau_K\otimes\rho_{V^NE_N'F_{RE}^{N_S}Y|\Omega})\\
    \leq p_{\Omega}\sum_{j=1}^{N_S}
    \Delta_t(\rho_{K_j\cdots K_{N_S}V^NE_N'F_{RE}^{N_S}Y|\Omega},\\
    \tau_{K_j}\otimes\rho_{K_{j+1}\cdots K_{N_S}V^NE_N'F_{RE}^{N_S}Y|\Omega}).
\end{multline}
Since sub-block hashing computes each output string $K_j=f_{RE,j}(A_{S_j}')$, one can apply the quantum leftover hash lemma~\cite{Tomamichel2011,Tomamichel2017} on each sub-block.
Therefore, one could set 
\begin{multline}
\label{Eqn:ErrorSamplingSum}
    \varepsilon'=p_{\Omega}\sum_{j=1}^{N_S}\{2\varepsilon_{sm,j}+\frac{1}{2}\\
    \times 2^{-\frac{1}{2}[H_{min}^{\varepsilon_{sm,j}}(A_{S_j}'|K_{j+1}\cdots K_{N_S}V^NE_N'F_{RE}^{N_S \setminus j}Y)-\tilde{l}]}\}
\end{multline}
to achieve the required security level, where the min-entropy is evaluated on $\rho_{A_{S_j}'K_{j+1}\cdots K_{N_S}V^NE_N'F^{N_S\setminus j}Y|\Omega}$, and $F_{RE}^{N_S\setminus j}$ refers to the set of hash functions used except $f_{RE,j}$.\\

To determine the error, the important quantity to compute is the min-entropy.
One can begin by lower bounding the min-entropy using the data-processing inequality, since $K_i=f_{RE,i}(A_{S_i}')$, and $A_{S_i}'$ can be computed from $A_{S_i}$,
\begin{equation}
\begin{split}
    &H_{min}^{\varepsilon_{sm,j}}(A_{S_j}'|K_{j+1}\cdots K_{N_S}V^NE_N'F_{RE}^{N_S\setminus j}Y)\\
    \geq& H_{min}^{\varepsilon_{sm,j}}(A_{S_j}'|A_{S_{j+1}}'\cdots A_{S_{N_S}}'V^NE_N'F^{N_S\setminus j}Y)\\
    \geq& H_{min}^{\varepsilon_{sm,j}}(A_{S_j}'|A_{S_{j+1}}\cdots A_{S_{N_S}}V^NE_N'Y),
\end{split}
\end{equation}
where the second inequality stems from the fact that $F_{RE}^{N_S\setminus j}$ is independent on all other terms in the min-entropy.\\

To analyse the min-entropy with generalised EAT, one has to remove the conditioning on $Y$, which, by definition, cannot be expressed as being generated round-by-round.
Depending on the nature of $Y$, there are many ways to remove the conditioning with min-entropy chain rule.
Here, we highlight one possible efficient method, if $Y$ is generated by blocks.
Suppose $Y_j$ is generated from the information in rounds $S_j$ only, i.e. $Y_i=g(A_{S_i}',I_{S_i})$, then one could remove the conditioning with
\begin{equation}
\begin{split}
    &H_{min}^{\varepsilon_{sm,j}}(A_{S_j}'|A_{S_{j+1}}'\cdots A_{S_{N_S}}'V^NE_N'Y)\\
    \geq& H_{min}^{\varepsilon_{sm,j}}(A_{S_j}'|A_{S^{\setminus j}}'V^NE_N'Y)\\
    \geq&H_{min}^{\varepsilon_{sm,j}}(A_{S_j}'|A_{S^{\setminus j}}'V^NE_N'Y_j)\\
    \geq& H_{min}^{\varepsilon_{sm,j}}(A_{S_j}'|A_{S^{\setminus j}}'V^NE_N')-\log_2\abs{\vars{Y}_j},
\end{split}
\end{equation}
where $S^{\setminus j}=\cup_{i:i\neq j}S_i$. 
The second inequality makes use of the property that conditioning does not increase min-entropy, the third inequality uses the data-processing inequality on $Y_i$ for $i\neq j$, since $A_{S_i}'$ and $I_{S_i}$ (in $E_N'$) are present, and the final inequality uses the chain rule.
For consistency, however, we shall use the min-entropy expansion in the main text in the following analysis (incurs a penalty $\log_2\abs{\vars{Y}}$) but it is straightforward to generalise the result to the more efficient case highlighted here.\\

We shall now focus on the min entropy term $H_{min}^{\varepsilon_{sm,j}}(A_j'|A_{S_j}'\cdots A_{S_{N_S}}'V^NE_N')$, which is now amenable to generalised EAT analysis.
In particular, one can define the channel $\tilde{\vars{M}}_i^j:R_{i-1}E_{i-1}\rightarrow R_iE_iA_{S_j,i}'C_i$, with $E_i=E_i''A_{S_{j+1,i}}'\cdots A_{S_{N_S,i}}'V_i$, and $E_i''=E_i'A_{S_{j+1}}^{'{i-1}}\cdots A_{S_{N_S}}^{'{i-1}}V^{i-1}$, 
\begin{enumerate}
    \item Implement EAT channel $\vars{M}_i$, which maps from $R_{i-1}E_{i-1}'$ to $R_iE_i'A_i'C_i$.
    \item Generate a uniformly random variable $V_i$ (step 1 of sample and hash).
    \item Define $A_{S_k,i}'=\begin{cases} A_i' & V_i=k \\ \perp & V_i\neq k\end{cases}$, for $k=j,\cdots,N_S$.
    \item Announce the values of $A_{S_k,i}'$ for $k=j+1,\cdots,N_S$ and $V_i$, i.e. they will be known to the adversary.
    \item Trace out $A_i'$.
\end{enumerate}
Applying this channel yields the state $\rho_{A_{S_j}'\cdots A_{S_{N_S}}'V^NC^NR_NE_N'}$ at the end of the protocol, with $E_N=V^NA_{S_{j+1}}'\cdots A_{S_{N_S}}'E_N'$ 
being known to the adversary.
To show that this is a valid EAT channel, one must check for the two conditions of projective reconstructability and non-signalling.\\

The statistics generated in $\{\tilde{\vars{M}}_i^j\}_i$, $C^N$, is no different from that in $\{\vars{M}_i\}_i$.
Moreover, since $C_i$ is generated from $(I^{test})^N$, which includes $A_i$ by assumption, the same generation method can be utilised in analysing $\tilde{\vars{M}}_i^j$.
Therefore, using $\vars{T}$ from the projective reconstructability of the $\vars{M}_i$ EAT channel, one can create $\vars{T'}$, which acts on the expanded $E_N=E_N'V^NA_{S_{j+1}}'\cdots A_{S_{N_S}}'$ with $\vars{T'}=\vars{T}\otimes\mathbb{I}_{V^NA_{S_{j+1}}'\cdots A_{S_{N_S}}'}$.
It is clear that $\vars{T'}\circ\Tr_{C_N}\tilde{\vars{M}}_N^j\circ\cdots\circ\Tr_{C_1}\circ\tilde{\vars{M}}_1^j=\tilde{\vars{M}}_N^j\circ\cdots\circ\tilde{\vars{M}}_1^j$, and thus $\{\tilde{\vars{M}}_N^j\}_i$ satisfies projective reconstructability.\\

The proof of non-signalling relies on the assumption that the original EAT channel $\vars{M}_i$ is non-signalling and that $A_i'$ is generated independently of any memory, i.e. $R_{i-1}$ of the EAT channel.
Due to this independence, one could in general split the EAT channel $\vars{M}_i$ into two parts.
The first part $\mathcal{E}_1:E_{i-1}'\rightarrow A_i'Q_i\tilde{E}_i$ 
is responsible for the generation of $A_i'$, taking only inputs $E_{i-1}'$ since it is independent on $A_i'$, with modifications to form $\tilde{E}_i$ and possible intermediate systems $Q_i$ generated.
The second part $\mathcal{E}_2:R_{i-1}\tilde{E}_iA_i'Q_i\rightarrow R_iE_i'A_i'C_i$ includes the use of $R_{i-1}$ to arrive at the final output systems of the $\vars{M}_i$ channel.
Since the first part does not depend on $R_{i-1}$, the no-signalling condition would only depend on the second part, i.e. that there exists $\tilde{R}_i':A_i'Q_i\tilde{E}_i\rightarrow E_i'$ such that $\Tr_{A_i'R_i}\circ\mathcal{E}_2\circ\mathcal{E}_1=\tilde{R}_i'\circ\mathcal{E}_1\circ\Tr_{R_{i-1}}$.\\

Consider the channel $\tilde{\vars{M}}_i^j$ described above, which can be written as a series of channels, $\tilde{\vars{M}}_i^j=\Tr_{A_i'}\circ\mathcal{E}_{A_i'\rightarrow V_iA_i'A_{S_{j,i}}'\cdots A_{S_{N_S,i}}'}\circ \vars{M}_i$, where $\mathcal{E}_{A_i'\rightarrow A_i'V_iA_{S_{j,i}}'\cdots A_{S_{N_S,i}}'}$ describes the generation of $V_i$ and the sampling step.
It can be noted that the order of the maps $\mathcal{E}_{A_i'\rightarrow A_i'V_iA_{S_{j,i}}'\cdots A_{S_{N_S,i}}'}$ and $\mathcal{E}_2$ are interchangeable, as neither act to alter the systems that are inputs to the other channel.
As such, we have that
\begin{equation}
\begin{split}
    &\Tr_{A_{S_j,i}'R_i}\circ\tilde{\vars{M}}_i^j\\
    =&\Tr_{A_{S_j,i}'R_i}\circ\Tr_{A_i'}\circ\mathcal{E}_{A_i'\rightarrow V_iA_i'A_{S_{j,i}}'\cdots A_{S_{N_S,i}}'}\circ \mathcal{E}_2\circ\mathcal{E}_1\\
    =&\Tr_{A_i'R_i}\circ\mathcal{E}_{A_i'\rightarrow V_iA_i'A_{S_{j+1,i}}'\cdots A_{S_{N_S,i}}'}\circ \mathcal{E}_2\circ\mathcal{E}_1\\
    =&\Tr_{A_i'R_i}\circ \mathcal{E}_2\circ\mathcal{E}_{A_i'\rightarrow V_iA_i'A_{S_{j+1,i}}'\cdots A_{S_{N_S,i}}'}\circ\mathcal{E}_1\\
    =&\tilde{R}_i'\circ\Tr_{R_{i-1}}\circ\mathcal{E}_{A_i'\rightarrow V_iA_{S_{j+1,i}}'\cdots A_{S_{N_S,i}}'}\circ\mathcal{E}_1\\
    =&\tilde{R}_i'\circ\mathcal{E}_{A_i'\rightarrow V_iA_{S_{j+1,i}}'\cdots A_{S_{N_S,i}}'}\circ\mathcal{E}_1\circ\Tr_{R_{i-1}},
\end{split}
\end{equation}
where the second equality traces out the output $A_{S_j,i}'$, the fourth equality applies the non-signalling property, and the final equality swaps the order of the trace over $R_{i-1}$ since neither $\mathcal{E}_1$ nor $\mathcal{E}_{A_i'\rightarrow V^iA_{S_{j+1,i}}'\cdots A_{S_{N_S,i}}'}$ act on $R_{i-1}$.
Noting that $\vars{R}_i=\tilde{R}_i'\circ\mathcal{E}_{A_i'\rightarrow V^iA_{S_{j+1,i}}'\cdots A_{S_{N_S,i}}'}\circ\mathcal{E}_1$ maps $E_{i-1}$ to $E_i=V_iE_i'A_{S_{j+1}}^{'i-1}\cdots A_{S_{N_S}}^{'i-1}A_{S_{j+1,i}}'\cdots A_{S_{N_S,i}}'$, the channel $\tilde{\vars{M}}_i^j$ is non-signalling.

With $\tilde{\vars{M}}_i^j$ as a valid EAT channel, one can now consider the min-tradeoff function to use before generalised EAT is applied.
In this case, the min-tradeoff function has to satisfy 
\begin{equation}
    f'(q)\leq\min_{\rho\in\Sigma_i'(q)} H(A_{S_j,i}'|A_{S_{j+1},i}'\cdots A_{S_{N_S},i}'V_iE_i''\tilde{E}_{i-1})_{\rho}
\end{equation}
with $\Sigma_i'(q)=\{\rho|\rho_{A_{S_j,i}'R_iE_i\tilde{E}_{i-1}}=\tilde{\vars{M}}_i^j(\omega_{R_{i-1}E_{i-1}\tilde{E}_{i-1}}),\rho_{C_i}=q\}$, where $E_i=E_i''A_{S_{j+1},i}'\cdots A_{S_{N_S},i}'V_i$.
The von Neumann entropy can be simplified by expanding in terms of different $v_i$ values. 
When $v_i\neq j$, $A_{S_j,i}'=\perp$, which has entropy 0 since it is a fixed value.
When $v_i=j$, $A_{S_j,i}'=A_i'$ and $A_{S_{k,i}}'=\perp$ for $k>j$.
Since $v_i$ is fixed as $j$, with $A_{S_{k,i}}'$ for $k>j$ containing no information on $A_i'$, the conditional entropy can be reduced to $H(A_i'|E_i''\tilde{E}_{i-1})$.
Since $A_{S_{j+1}^{'i-1}}\cdots A_{S_{N_S}^{'i-1}}V^{i-1}$ are not used in channel $\tilde{\vars{M}}_i^j$, they do not give any information on $A_i'$ and we can reduce $H(A_i'|E_i''\tilde{E}_{i-1})=H(A_i'|E_i'\tilde{E}_{i-1})$.
Moreover, since the generation of $A_i'$ and $E_i'$ in $\tilde{\vars{M}}_i^j$ are fully determined by the channel $\vars{M}_i$, the states to optimise over reduced to $\Sigma_i(q)$, thus yielding the requirement on the min-tradeoff function
\begin{equation}
    f'(q)\leq \min_{\rho\in\Sigma_i(q)} p_S H(A_i'|E_i'\tilde{E}_{i-1})_{\rho},
\end{equation}
where the $p_S$ factor is due to the probability of $v_i=j$.
From the GEAT-analysable original protocol, there exists a min-tradeoff function $f(q)$ that lower bounds the RHS without the factor $p_S$.
Therefore, $f'(q)=p_Sf(q)$ is a valid min-tradeoff function.\\

Applying GEAT using this min-tradeoff function, along with EAT channels $\tilde{\vars{M}}_i^j$, one arrives at the result in the main text. Since this result applies to all min-entropy terms in Eqn.~\ref{Eqn:ErrorSamplingSum}, we have that
\begin{multline}
    \varepsilon'=p_{\Omega}\sum_{j=1}^{N_S}\{2\varepsilon_{sm,j}+\frac{1}{2}\times \\
    2^{-\frac{1}{2}[Nh'-v_1'\sqrt{N}-v_0'-\log_2\abs{\vars{Y}}-\tilde{l}]}\}.
\end{multline}
If the smoothing parameters are set to be equal for all rounds, $\varepsilon_{sm,j}=\tilde{\varepsilon}_{sm}$, the error reduces to 
\begin{equation}
    \varepsilon'=N_Sp_{\Omega}[2\tilde{\varepsilon}_{sm}+\frac{1}{2}\times \\
    2^{-\frac{1}{2}[Nh'-v_1'\sqrt{N}-v_0'-\log_2\abs{\vars{Y}}-\tilde{l}]}].
\end{equation}

\section{BBM92 Security Analysis}
\label{App:BB84Sec}

In this section, we provide a security analysis with the BBM92 protocol, which is utilised as an example to illustrate the effects of our proposed sub-block sampling method. 
We consider here a simple example of BBM92, where an untrusted third party, Eve, sends $N$ pairs of qubits to Alice and Bob. Alice and Bob use $\sZ$ basis measurements as key generation basis and $\sX$ basis to check for Eve's attack, with the following protocol:
\begin{enumerate}
    \item For each round, Eve prepares entangled qubit state $\rho_{Q_{A,i}Q_{B,i}E}$, and sends subsystem $Q_{A,i}$ to Alice and $Q_{B,i}$ to Bob.
    \item Alice (resp. Bob) randomly chooses a basis $D_i\in\{\sX, \sZ\}$ (resp. $D'_i$), with probabilities $p_{\sX}$ and $p_{\sZ}$, and measures their part of the state in the chosen basis to obtain outcome $A_i$ (resp. $B_i$). If Alice (resp. Bob) obtains a no detection outcome, then she (he) sets $A_i = \perp$, ($B_i = \perp$). When Alice (resp. Bob) obtains double detection, she (he) randomly assigns a bit value with a uniform probability to $A_i$ (resp. $B_i$).
    \item The quantum state preparation and measurement are repeated for $N$ rounds.
    \item For each round, Alice and Bob announce their basis choice, $D_i$ and $D'_i$. Furthermore, they announce $W_i = \begin{cases}1 & \text{if } A_i = \perp \\
    0 & \text{otherwise} \end{cases}$ and $W'_i = \begin{cases}1 & \text{if } B_i = \perp \\
    0 & \text{otherwise} \end{cases}$. If either $W_i = 1$ or $W'_i = 1$, both parties set their outputs to $A_i = B_i = \perp$.
    \item Alice and Bob compute the sifted key $A_i'=\begin{cases} A_i, & D_i=D_i'= \sZ\\ \perp, & \text{otherwise} \end{cases}$ and $B_i'=\begin{cases} B_i, & D_i=D_i'= \sZ\\ \perp, & \text{otherwise} \end{cases}$ to include rounds where both parties select the $\sZ$ basis.
    \item Alice and Bob label the test round with $T_i=1$, when they both choose the $\sX$ basis for measurement, i.e. $T_i=\begin{cases} 1, & D_i=D_i'= \sX\\ 0, & \text{otherwise} \end{cases}$. Alice and Bob announce the test round measurement outcomes, $A_i^{test}=\begin{cases} A_i & T_i=1\\ \perp & T_i=0 \end{cases}$ and $B_i^{test}=\begin{cases} B_i & T_i=1\\ \perp & T_i=0 \end{cases}$.
    \item Using the revealed test round outcomes, both parties generate the statistics $C_i=\begin{cases} 0 & T_i=1,\,W_i=W_i'=0,\,A_i^{test}=B_i^{test}\\ 1 & T_i=1,\,W_i=W_i'=0,\,A_i^{test}\neq B_i^{test}\\ 2 & T_i=1,\,W_i\neq 0\, \text{ or }\, W_i'\neq 0\\ \perp & T_i=0 \end{cases}$, where $c=\perp$ indicates that the round is not a test round, $c=0$ represents the correct measurement outcomes, and $c=1$ represents the rounds where the outcomes are mismatched and $c=2$ 
    recorded rounds in the $\sX$ basis where no detection occurs. 
    \item Based on the computed statistics, the users abort the protocol if the number of error bits exceed some threshold $\text{freq}(c^N)_1>p_{\sX}^2 \eta_{tol} Q_{tol}$, where $Q_{tol}$ is the tolerated average error rate and $\eta_{tol}$ is the tolerated minimum transmission, or if the number of no detection rounds are large, $\text{freq}(c^N)_2>p_{\sX}^2(1-\eta_{tol})$.
    \item Alice and Bob perform error correction, with the overall syndrome communicated, $V_{EC}$, being $l_{EC}$ in length.
    \item Alice and Bob perform privacy amplification to obtain the final secure key $K=F(A^{'N})$ and $\hat{K}=F(B^{'N})$.
\end{enumerate}

In the protocol, Eve's most general attack is to prepare some quantum state $\rho_{Q_A^NQ_B^NE}$, and send $Q_{A,i}$ to Alice and $Q_{B,i}$ to Bob in each round.
Eve also has access to any publicly communicated information, and thus her side information is $F_{RE}E_N'V_{EC}$, where $E_N'=EW^NW^{'N}I^N(I^{test})^N$, $I^N=(D^N,D^{'N})$,  $(I^{test})^N=[T^N,(A^{test})^N,(B^{test})^N]$.
For simplicity, we focus only on the secrecy condition of QKD,
\begin{equation}
    p_{\Omega}\Delta(\rho_{KE_N'F_{RE}|\Omega},\tau_K\otimes \rho_{E_N'F_{RE}|\Omega})\leq\varepsilon_{sec}.
\end{equation}
Using the quantum leftover hash lemma~\cite{Tomamichel2011,Tomamichel2017}, the error can be expressed as
\begin{equation}
    \varepsilon_{sec}=2\varepsilon_{sm}+\frac{1}{2}\times 2^{-\frac{1}{2}[H_{min}^{\varepsilon_{sm}}(A^{'N}|E_N'V_{EC})-l]},
\end{equation}
where the min-entropy is computed for $\rho_{A^{'N}E_N'V_{EC}|\Omega}$.
Since $V_{EC}$ is not generated in a round-by-round manner ($Y$ in generic GEAT-analysable protocol), it is removed using the chain rule of smooth min-entropy~\cite{Vitanov2013,Tomamichel2016},
\begin{equation}
    H_{min}^{\varepsilon_{sm}}(A^{'N}|E_N'V_{EC})\geq H_{min}^{\varepsilon_{sm}}(A^{'N}|E_N')-l_{EC},
\end{equation}
where for simplicity, we assume $l_{EC}=Nf_{EC}h_b(e_b)$, where $e_b$ is the bit error rate, and $f_{EC}$ is the efficiency of the error correction step.\\

We now focus on simplifying $H_{min}^{\varepsilon_{sm}}(A^{'N}|E_N')$ using generalised EAT.
In the most general case, Eve's attack can be performed at the start, preparing the quantum state $\rho_{Q_A^NQ_B^NE}$, before sending the sub-system to Alice and Bob.
Therefore, one can craft the EAT channel, with initial state $\rho_{E_0'}=\rho_{Q_A^NQ_B^NE}$ sent into EAT channel, $\vars{M}_i:E_{i-1}'\rightarrow E_i'A_i'C_i$, where $E_i'=Q_{A,i+1}^NQ_{B,i+1}^NEI^i(I^{test})^i$, with $Q_{A,i+1}^NQ_{B,i+1}^N=\{Q_{A,j}Q_{B,j}\}_{j=i+1,\cdots,N}$.
To achieve the same statistics, $\rho_{A^{'N}E_N'C^N}=\vars{M}_N\circ\cdots\circ\vars{M}_1(\rho_{Q_A^NQ_B^NE})$, one can formulate the EAT channel as follows:
\begin{enumerate}
    \item Randomly select $D_i\in\{\sX,\sZ\}$ and $D_i'$ with probabilities $p_{\sX}$ and $p_{\sZ}$.
    \item Measure subsystems $Q_{A,i}$ and $Q_{B,i}$ in the chosen basis and obtain outcomes $A_i$ and $B_i$.
    \item Compute the detection result, $W_i$, $W_i'$, test round choice $T_i$, test round outcomes $A_i^{test}$, $B_i^{test}$, sifted bits $A_i'$, $B_i'$ and statistics $C_i$.
    \item Trace out $A_i$, $B_i$ and $B_i'$. 
\end{enumerate}
Since $T_i$, $A_i^{test}$ and $B_i^{test}$ is announced by the two parties, they are accessible to Eve.
Therefore, it is clear that $C^N$ can be recovered through $E_N'$ alone and the channel satisfies projective reconstructability.
Since there is no $R_i$ in the protocol, non-signalling is trivially satisfied by $\vars{R}_i=\Tr_{A_i}\circ\vars{M}_i$, demonstrating that $\vars{M}_i$ is a valid EAT channel.\\

Before applying GEAT, one has to define a valid min-tradeoff function, $f(q)$, which lower bounds $H(A_i'|E_i')$ for any state that can be generated from the channel $\vars{M}_i$ and has statistics $C_i\in\{0,1,2,\perp\}$ satisfying some probability distribution $q=(q_0,q_1,q_2,q_{\perp})$, i.e. states in set $\Sigma_i(q)=\{\rho|\rho_{A_i'C_iR_iE_i\tilde{E}_{i-1}}=\vars{M}_i(\omega_{R_{i-1}E_{i-1}\tilde{E}_{i-1}}),\rho_{C_i}=q\}$.
Expanding the entropy to remove cases where $A_i'=\perp$, when Alice and Bob did not both choose the $\sZ$ basis or when either party registers no detection,
\begin{equation}
    H(A_i'|E_i'\tilde{E}_{i-1})=p_{\sZ}^2p_{det}H(A_i^{det,\sZ \sZ}|E\tilde{E}_{i-1}),
\end{equation}
where $A_i^{det,\sZ \sZ}$ are the outcomes of rounds where both parties chose a $\sZ$ basis and records a detection and $p_{det}$ is the detection probability.
Note that fair sampling is assumed here, $p_{det|\sX \sX}=p_{det|\sZ \sZ}=p_{det}$, so the detection probability should be basis independent.
Since these rounds have guaranteed detection and $\sZ$ basis measurement, one can apply the entropic uncertainty relation 
 (for conditional von Neumann entropy)~\cite{Coles2017,berta2010uncertainty} to show
\begin{equation}
    H(A_i'|E_i'\tilde{E}_{i-1})\geq p_{\sZ}^2p_{det} [1-h_{b}(e_{ph})],
\end{equation}
where the phase error rate can be computed from $e_{ph}=\frac{q_1}{q_1+q_0}$ and the detection probability from $p_{det}=(q_0+q_1) / p_{\sX}^2=1-\frac{q_2}{p_{\sX}^2}$ for statistics $\vec{q}$.
To define an affine min-tradeoff function, noting that binary entropy is concave, one could upper bound the expression above by its tangent, 
\begin{multline}
    H(A_i'|E_i'\tilde{E}_{i-1})\geq \\
    p_{\sZ}^2p_{det} \left[ 1-h_{b}(e_{ph}')-\log_2\left(\frac{1}{e_{ph}'}-1 \right) \left(\frac{q_1}{p_{\sX}^2  p_{det}}-e_{ph}' \right) \right],
\end{multline}
where $e_{ph}'\in (0,0.5)$ is the tangent point, which could be optimised to provide a tight entropy bound.
Simplifying the RHS of the equation, one arrives at an affine min-tradeoff function
\begin{multline}
    f(\vec{q})=p_{\sZ}^2 \Bigg\{\left(1-\frac{q_2}{p_{\sX}^2}\right) \left[1+\log_2(1-e_{ph}') \right]\\
    -\frac{q_1}{p_{\sX}^2}\log_2\left(\frac{1}{e_{ph}'}-1\right) \Bigg\}.
\end{multline}
Using the min-tradeoff function, one could compute the quantities relevant to GEAT,
\begin{gather*}
    \mathrm{Max}(f)=p_{\sZ}^2 \left[1+\log_2(1-e_{ph}') \right]\\
    \mathrm{Min}_{\Sigma}(f)= p_{\sZ}^2\left[1+\log_2 e_{ph}'\right]\\
    \mathrm{Var}(f)\leq\frac{p_{\sZ}^4}{p_{\sX}^2} \left[1+\log_2(1-e_{ph}') \right]^2\\
    h=p_{\sZ}^2\eta_{tol} \left[1-h(e_{ph}') - (Q_{tol} - e_{ph}') \log_2\left(\frac{1}{e_{ph}'}-1\right) \right]
\end{gather*}
using an expansion similar to those in Lemma V.5 of Ref.~\cite{Dupuis2019} for the variance computation.
Applying GEAT, the error can now be expressed as
\begin{multline}
    \varepsilon_{sec}=2\varepsilon_{sm}+\frac{1}{2}\times 2^{-\frac{1}{2}[Nh-Nf_{EC}h_b(e_b)-l]}\\
    \times 2^{-\frac{1}{2}[-N\frac{(\alpha-1)\ln 2}{2(2-\alpha)}V^2-N(\frac{\alpha-1}{2-\alpha})^2K'(\alpha)-\frac{g(\varepsilon)+\alpha\log_2(\frac{1}{p_{\Omega}})}{\alpha-1}]}.
\end{multline}
The key length at a given secrecy level can then be expressed as
\begin{multline}
    l\geq Nh-N\frac{(\alpha-1)\ln 2}{2(2-\alpha)}V^2-N\left(\frac{\alpha-1}{2-\alpha}\right)^2K'(\alpha)\\
    -\frac{g(\varepsilon_{sm})+\alpha\log_2(\frac{1}{p_{\Omega}})}{\alpha-1}-N p_{\mathsf{Z}}^2 p_{det} f_{EC}h_b(e_b)\\+1-\log_2\left(\frac{1}{\varepsilon_{sec}-2\varepsilon_{sm}}\right).
\end{multline}

\end{document}